%% file: Coma_dark_matter_nov.tex
\documentclass[twocolumn]{emulateapj}

\usepackage{comment}
\usepackage{color}

\newcommand\deltaphi{\delta\Phi}
\newcommand\mh{m_{\rm H}}
\newcommand\mtot{M_{\rm t}}
\newcommand\rc{R_{\rm c}}
\newcommand\rhoscale{\rho_0}
\newcommand\rs{a}
\newcommand\rt{r_{\rm t}}
\newcommand\zmax{z_{\rm max}}
\newcommand\rp{\varpi_{\rm p}}
\newcommand\zp{z_{\rm p}}
\newcommand\rvec{\mathbf{r}}
\newcommand\rveci{\rvec_{\rm i}}
\newcommand\vvec{\mathbf{v}}
\newcommand\ie{i.e.}
\newcommand\eg{e.g.}
\newcommand\rhobar{\overline{\rho}}
\newcommand\rhot{\rhobar_{\rm t}}
\newcommand\rhovir{\rhobar_{\rm vir}}
\newcommand\rvir{r_{\rm vir}}
\newcommand\xmm{{\it XMM-Newton}}

\newcommand\dof{{\rm dof}}
\newcommand\msun{M_\odot}
\newcommand\chandra{{\it Chandra}}
\newcommand\kb{k_{\rm B}}
\newcommand\vhat{\widehat\vvec_0}
\newcommand\rhat{\widehat\rvec_0(t)}
\newcommand\wt{w_{\rm t}}

\shorttitle{Dark matter subhalos in Coma}
\shortauthors{Andrade-Santos et al.}

\begin{document}


\title{Dark Matter Subhalos and the X-ray Morphology of the Coma Cluster}



\author{
Felipe Andrade-Santos\altaffilmark{1,2}, 
Paul E. J. Nulsen\altaffilmark{1},
Ralph P. Kraft\altaffilmark{1},
William R. Forman\altaffilmark{1},\\
Christine Jones\altaffilmark{1},
Eugene Churazov\altaffilmark{3}, and
Alexey Vikhlinin\altaffilmark{1}
}
\affil{$^1$Harvard-Smithsonian Center for Astrophysics, 60 Garden
  Street, Cambridge, MA 02138, USA}   
\affil{$^2$Departamento de Astronomia - IAG, Universidade de S\~{a}o Paulo,
    S\~{a}o Paulo - SP, Brazil}
\affil{$^3$Max Planck Institute for Astrophysics, Karl-Schwarzschild
  str. 1, Garching D-85741, Germany}




\begin{abstract}
Structure formation models predict that clusters of galaxies contain
numerous massive subhalos.  The gravity of a subhalo 
in a cluster
compresses the surrounding intracluster gas and enhances its X-ray
emission.  We present a simple model, which treats subhalos as slow
moving and gasless, for computing this effect.  Recent weak lensing
measurements by Okabe et al. have determined masses of
$\sim 10^{13}~\msun$ for three mass concentrations projected within 300
kpc of the center of the Coma Cluster, two of which are centered on
the giant elliptical galaxies NGC~4889 and NGC~4874.  Adopting a
smooth spheroidal $\beta$-model for the gas distribution in the
unperturbed cluster, we model the effect of these subhalos on the X-ray
morphology of the Coma Cluster, comparing our results to \chandra{}
and \xmm{} X-ray data.  The agreement between the models
and the X-ray
morphology of the central Coma Cluster is striking.  With subhalo
parameters from the lensing measurements, the distances 
of the three subhalos from the Coma
Cluster midplane along our line of sight  are
all tightly constrained.  Using the model to fit the subhalo masses
for NGC~4889 and NGC~4874 gives $9.1 \times 10^{12}~\msun$ and $7.6
\times 10^{12}~\msun$, respectively, in good agreement with the
lensing masses.  These results lend strong support to the argument
that NGC~4889 and NGC~4874 are each associated with a subhalo that
resides near the center of the Coma Cluster.  In addition to
constraining the masses and 3-d location of subhalos, the X-ray data
show promise as a means of probing the structure of central subhalos.
\end{abstract}



\section{Introduction}

In the standard $\Lambda$CDM cosmology, the massive halos dominated by
dark matter that host galaxies, groups, and clusters of galaxies
assemble by mergers of smaller structures.  Under the influence of
gravity, uncollapsed matter and smaller collapsed halos fall into
larger halos and, occasionally, halos of comparable mass merge with
one another.  Earlier generations of dark matter halos can survive as
subhalos within the resulting aggregates, causing significant
departures from the idealized smooth mass distributions often used to
model dark matter halos.  Hierarchical collapse models make clear
predictions for the level of substructure within dark matter halos,
which can provide tests of the $\Lambda$CDM model.  A possible
conflict with these predictions for galaxy scale dark halos is the
basis of the ``missing dwarf problem'' \citep{kkv99, mgg99, rbk12}.

Numerical simulations of structure formation find that many ($\sim
10$) dark matter halos with masses of $\sim 10^{12}~\msun$ would
be accreted by a rich galaxy cluster during the past few Gyr.  Using
the Millennium I $\&$ II simulations \citep{2005Springel, 2009Boylan},
\citet{2010Fak} find that the Coma Cluster ($M_{\rm tot} = 2 \times
10^{15}~\msun$) would have accreted $\sim 7$ dark matter halos
with masses in the range $10^{12}$ -- $10^{13}~\msun$ in the last Gyr.
Supporting this prediction, the analysis of \citet{2010Giocoli}, which
identifies substructures at all levels of the hierarchy (subhalos,
sub-subhalos, etc.), finds that there should be $\sim 7$ dark matter
substructures, with masses in this range, present in a cluster like Coma
today.

Recently, \citet{cvz12}  analyzed X-ray surface brightness
fluctuations in the central region of the Coma Cluster.  They discuss
the physical origin of these fluctuations, noting in particular that
fluctuations in the gravitational potential associated with the two
massive central galaxies in Coma could account for some of the larger
scale surface brightness fluctuations in that region \citep[see
  also][]{1994Vik}.  In this paper, we present a more detailed model
for the effect of subhalos, quantified by gravitational lensing
measurements, on the X-ray surface brightness in the central region of
the Coma Cluster.

\subsection{The Coma Cluster} \label{sec:coma}

The Coma Cluster ($z = 0.0236$) has long been considered typical of
dynamically relaxed systems.  However, a wide range of studies have
revealed that it contains complex substructure.  Optical signatures of
substructure come from the work of \citet{1987Fitchett}, who used a
maximum likelihood method to partition the cluster into subsystems
that have yet to come into dynamical equilibrium with one another, and
\citet{1988Mellier}, who determined mass-to-light ratios for the
substructure surrounding the two dominant galaxies in the cluster
core.  \citet{1996Colless} used a sample of 552 galaxy redshifts to
further clarify the merger history of the Coma Cluster.  X-ray studies
also reveal a remarkable complexity. \citet{1992Briel} detected
  diffuse X-ray emission from the regions of NGC 4839 and 4911 subgroups
at 6\% and 1\% of the total cluster emission, respectively.
\citet{1993White} showed that the Coma Cluster was formed by 
the merging of several distinct subunits which are not yet fully
destroyed
 and \citet{1994Vik} showed that the extended regions of X-ray
 emission in the central region of Coma are associated with
 the subgroups NGC 4889 and 4874.  \citet{1996Biviano} combined {\it ROSAT} and
extensive optical redshift and photometric samples to identify the
main body of the cluster, suggesting that it is rotating.  They also
concluded, based on the differences in velocity between the NGC~4889
and NGC~4874 groups and the cluster mean, that these groups only
arrived recently in the cluster core. 
In a more recent work, \citet{2007A&A...468..815G} used planetary nebulae
as tracers to investigate the ongoing subcluster merger in the
Coma core. From the planetary nebulae line of sight velocity 
distribution they concluded that  the NGC 4889 subcluster is likely 
to have fallen into Coma from the eastern  A2199 filament, in a 
direction nearly in the plane of the sky, thus colliding with the
NGC 4874 subcluster coming from the west.
 \citet{2005Adami} presented a
multi-wavelength analysis of subclustering in Coma, using X-ray data
and a compilation of nearly 900 galaxy redshifts to partition the
cluster into 17 subgroups that they identify as the remnants of infalling
groups from the surrounding large-scale structure.

Further evidence of ongoing infall in Coma can be found in X-ray observations.
For example, the linear filament to the southeast 
identified by \citet{1997Vik} in the
Coma Cluster, which extends $\simeq 1$ Mpc from the cluster center
toward NGC~4911 and NGC~4921, may be enhanced X-ray emission due to
gas stripped from an infalling group or the potential perturbations
caused by tidally stripped dark matter.  In either case, it is
transient on the dynamical time scale and so must be due to recent
infall.  In the radio, \citet{br11} have argued that a $\sim2$ Mpc
radio relic on the western side of the cluster is one signature of
an infall shock, caused by a new, ongoing merger.  If so, this merger
is distinct from the one responsible for the formation of the X-ray
filament.

X-ray studies of galaxy clusters provide clues to the dynamical
effects of local matter concentrations on the intracluster (ICM).  The hot
intracluster plasma, which contains most of the baryonic mass in a
rich cluster like Coma, is perturbed by infalling groups.  Regions of
enhanced X-ray emission define dense gas concentrations, for which
there are a number of possible causes, including the
gravitational effect of dark matter subhalos on the hot, diffuse gas.
As mentioned, numerical simulations suggest that in the last Gyr of
its history, the Coma Cluster has accreted $\sim 7$ dark matter halos
with masses in the range $\sim 10^{12}$ -- $10^{13}~\msun$, which might be
detectable via their gravitational effect on the cluster gas.  Such an
effect has already been detected \citep{1997Vik}.  However, a
large-scale perturbation to the X-ray morphology has not previously
been associated directly with embedded dark matter subhalos.  Studies
of the effect of subhalos on the gas may help us to better understand
the cluster X-ray morphology and test predictions for the distribution
of dark matter subhalos in clusters.

\citet{2010Okabe} have measured the projected mass distribution within
$60'$ of the center of the Coma Cluster (around NGC 4874), using weak
lensing data from two Subaru/Suprime-Cam fields.  They detected eight
subclump candidates, with a mean projected mass for
seven\footnote{They quantified the contribution of background
  large-scale structure (LSS) to the projected mass distribution using
  Sloan Digital Sky Survey multi-band and photometric data, assuming a
  mass-to-light ratio appropriate for field galaxies.  They found that
  one of the eight subclump candidates, which is not associated with
  any member galaxies, is significantly affected by LSS lensing.}  of
the subclumps within the cluster equal to $7.2 \pm 1.9 \times
10^{12}~\msun$.  In principle, analysis of the X-ray surface
brightness of Coma can constrain the location of a dark matter
subclump along our line of sight within the cluster.  The gravity of a
subhalo compresses the intracluster gas and produces an enhancement
in X-ray brightness that depends on the density of the intracluster gas
that surrounds the subhalo.  Subhalos closer to the cluster center
will cause greater brightness enhancements simply because they are
embedded in denser gas.  To explore this possibility, we have modeled
enhancements in the X-ray surface brightness resulting from adiabatic
compression of the intracluster gas caused by dark matter subhalos
residing in an otherwise relaxed cluster potential.

In the following sections we present the observations of Coma used
here (Section \ref{sec:data}) and our model for estimating the
enhancement in X-ray surface brightness caused by a dark matter
subhalo embedded in the ICM (Section \ref{sec:model}).  In Section
\ref{sec:Num_approach} we present the results of numerical
calculations for three subhalos identified by \citet{2010Okabe}, followed
by a discussion in Section \ref{sec:discuss} and summary in Section
\ref{sec:conc}.  The cosmology assumed here has $\Omega_{\rm M}=0.3$,
$\Omega_{\Lambda}=0.7$ and $H_0=70$~km~s$^{-1}$Mpc$^{-1}$, implying a
linear scale of $0.48\rm\ kpc\ arcsec^{-1}$ at the Coma distance of
100 Mpc.


\section{X-ray Observations and Data Reduction} \label{sec:data}

\chandra{} and \xmm{} observations of the Coma Cluster were used make
maps of the hardness-ratio and X-ray surface brightness for comparison
with the models that are discussed in detail in later sections of this
paper.  In this section we outline the observations analysis.

\subsection{\chandra{} Observations}

The core of the Coma Cluster (A1656) has been observed several times by
\chandra, for a total exposure time on the ACIS-I and S detectors of
120 ks (ObsIDs
\dataset[ADS/Sa.CXO#obs/00555]{555},
\dataset[ADS/Sa.CXO#obs/00556]{556},
\dataset[ADS/Sa.CXO#obs/01086]{1086},
\dataset[ADS/Sa.CXO#obs/01112]{1112},
\dataset[ADS/Sa.CXO#obs/01113]{1113},
\dataset[ADS/Sa.CXO#obs/01114]{1114},
\dataset[ADS/Sa.CXO#obs/09714]{9714},
\dataset[ADS/Sa.CXO#obs/10672]{10672}).  
The data were reduced following standard CIAO 4.4 procedures, with
CALDB 4.5.3, 
including corrections for time dependence of the charge
transfer inefficiency and gain.  Data were also filtered for periods
of high background and standard blank sky background files were used
to obtain background images.\footnote{We note that the background
  level for the X-ray bright Coma Cluster is practically negligible.}
Events from individual observations were reprojected and co-added to
generate combined \chandra{} images with a total exposure of 120 ks.

\subsection{\xmm{} Observations}

From \xmm, we used the same 0.5--2.5 keV EPIC/MOS imaging data as
used by \citet{cvz12}.  In brief, \xmm{} pointings covering a field of
more than $1^\circ \times 1^\circ$ were combined into a single image,
the central part of which was fitted to our models.  Periods of
background flaring were filtered out based on the event rate above 10
keV.  Background images were made by scaling ``blank sky fields'' to
match observed count rates in the 11--12 keV band.  The image area
extends well outside the field covered by \chandra{}, to the outskirts
of the Coma Cluster.  The cleaned and calibrated MOS data were
combined to make a raw 0.5--2.5 keV image, together with a
background image and exposure map.  An error map was computed assuming
Poisson statistics for the image and background data.  Simulated
images are multiplied by the exposure map for fitting to the
background subtracted raw image.



\section{Subhalos in Coma}  \label{sec:model}

\begin{figure*}[!htb]
\begin{center}
\includegraphics[width=1.0\textwidth]{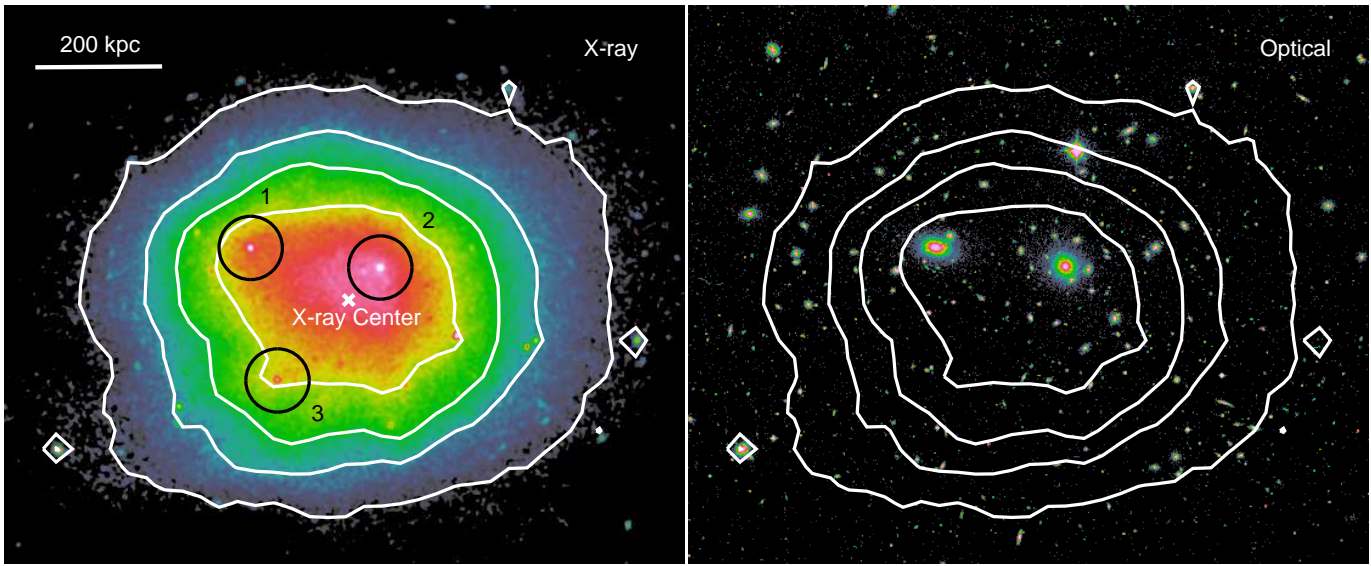}
\end{center}
\caption{Left: positions of the three central subhalos.  Subhalo 1 is
  centered on NGC~4889 (denoted by the number 1), subhalo 2 is
  centered on NGC~4874 (2), and subhalo 3, which has no optical
  counterpart, is centered on X-ray bright source closest to the
  lensing position from \citet{2010Okabe}.  
Right: the \xmm{} image contours are shown
  overlaid on the optical image (SAO-DSS) in the right panel.}
\label{clumps_okabe}
\end{figure*}

As discussed in Section \ref{sec:coma}, simulations of hierarchical
structure formation suggest that about seven dark matter subhalos with
masses of $\gtrsim 5 \times 10^{12}~\msun$ are expected to be present
in the Coma Cluster now.  Here we explore the observable impact of
dark matter subhalos on gas near the cluster center, where their
effect on the cluster X-ray emission is greatest.  Using lensing data,
\citet{2010Okabe} have reported the discovery of seven subclumps
with masses of $\sim 10^{13}~\msun$, three of which are marked in
Figure \ref{clumps_okabe}.  Two of these lie close to the giant
elliptical galaxies NGC~4874 and NGC~4889, and their masses are
consistent with the masses attributed to subgroups associated with
these two galaxies in previous studies \citep[subclumps 2 and 1,
  respectively, in Figure \ref{clumps_okabe};][]{1988Mellier,1994Vik}.
We begin by outlining a simple model for estimating the effect of
subhalos on the X-ray surface brightness of a cluster, which we then
apply to subhalos in the Coma Cluster.

\subsection{Impact on the Intracluster Gas} \label{sec:isent}

We  treat the subhalos as static perturbations on the large scale
gravitational potential of the cluster.  An off-center subhalo cannot
remain at rest, but, unless its speed is comparable to the sound speed
or greater, the effect of its motion on the gas is modest \citep[e.g.,
  see][]{2005Mach}.  The differences between the mean radial velocity
of the Coma Cluster and the radial velocities of NGC 4874 and NGC 4889
are $\Delta v_r$ = 299 and $-430\rm\ km\ s^{-1}$, respectively, both
small compared to the gas sound speed of $\simeq 1400\rm\ km\ s^{-1}$.
Although the transverse velocities of these galaxies are unknown, the
lack of evidence for shock fronts or substantial asymmetries in the
distribution of gas in their vicinities suggests that their motion
through the cluster is significantly subsonic.  Thus, the hydrostatic
approximation is likely to be valid for the associated subhalos.  More
generally, the motion of typical subhalos is expected to be roughly
transonic.  Although transonic motion certainly alters details of
perturbations to the gas density, the hydrostatic approximation will
provide a good estimate for the magnitude of the change in the X-ray
surface brightness, unless the speed of a subhalo is appreciably
supersonic (see the Appendix).

The effect of a static subhalo is to deepen the gravitational
potential locally, drawing in and compressing the gas, thereby
increasing the local X-ray emission.  At the relevant temperatures and for
typical element abundances, the broad band responses of both \chandra{}
and \xmm{} to optically thin thermal emission from hot gas are relatively
insensitive to the gas temperature.  For example, for a fixed emission
measure, the
0.5--2.5 keV count rate of the \chandra{} ACIS-I 
declines by a total of
$\simeq 17\%$ as the gas temperature increases from $\kb T = 4$ to
12 keV.  The temperature dependence of the \xmm{} response is similar.
Thus, we can reasonably ignore the temperature dependence of the
\chandra{} and \xmm{} responses and assume that the X-ray count rate
per unit volume of the gas depends only on its density, being directly
proportional to the square of the gas density.

Apart from tiny cores of cool gas in NGC 4889 and NGC 4874
\citep{2001Vik}, the cooling time of the gas near the center of the
Coma Cluster exceeds the Hubble time, so that the effects of radiative
cooling are negligible.  For a slowly moving subhalo, the response of
the gas is therefore adiabatic.  We assume that the perturbed
region is small compared to the scale over which gas properties vary
appreciably, so that the gas affected by a subhalo is approximately
isentropic.  The momentum equation for isentropic gas may be written 
\begin{equation}
{d\mathbf{v} \over dt} = - \nabla (H + \Phi),
\label{eqn:mom}
\end{equation}
where $\Phi$ is the gravitational potential, $\vvec$ is the gas
velocity, $H = 5 \kb T / (2 \mu \mh)$ is the specific enthalpy of the
gas, $T$ is its temperature, $\mu \mh$ is the mean mass per gas
particle, $\kb$ is the Boltzmann constant, and the ratio of specific
heats is assumed to be $\gamma = 5/3$  throughout this paper.
For slow motion, we can ignore $d \mathbf{v} / dt$ to obtain
\begin{equation}
H + \Phi \simeq \textrm{constant}
\label{eqn:pert}
\end{equation}
at any given time.  As discussed in the Appendix,
Equation (\ref{eqn:pert}) provides a good estimate for the local
perturbation due to a subhalo that moves at no more than transonic
speeds and has a small mass ($G \mtot / H_0 \ll \rc$, the cluster core
radius, in the notation below).  The subhalo also needs to have been
well stripped of its gas, since we ignore any possible effects due to
the motion through the ICM of a remaining atmosphere bound to the
subhalo.  This is generally a good approximation for subhalos near the
center of the Coma Cluster.  Note that our assumption that the
perturbations are adiabatic (isentropic) differs from the assumption
of isothermality employed by \citet{cvz12}.  Isentropic gas is less
compressible than isothermal gas, so that a dark halo will cause a
smaller change in surface brightness under the model used here (for
the same perturbation, $\delta\Phi$, expanded to first-order, the
change in surface brightness estimated here is 60\% of that of
\citealt{cvz12}).

Expressing the gravitational potential at $\rvec$ as $\Phi (\rvec) =
\Phi_0 (\rvec) + \deltaphi (\rvec)$, where $\Phi_0$ is the potential
of the unperturbed cluster and $\deltaphi$ is the potential of a
subhalo, the local effect of the potential perturbation is to modify
the specific enthalpy to $H \simeq H_0 - \deltaphi$, where $H_0$ is
the specific enthalpy of the unperturbed gas at the location of
interest.  For locally isentropic gas with $\gamma = 5/3$, the
perturbed electron density  is
\begin{equation}
n = n_0 \left(1 - \frac{\deltaphi}{H_0}\right)^{3/2},
\label{density_perturbation}
\end{equation}
where $n_0$ is the unperturbed electron density at the position of
interest.

\subsection{Conduction and the Adiabatic
  Approximation}  \label{sec:cond}

High thermal conductivity in the Coma ICM has the potential to make
the perturbations more nearly isothermal than adiabatic.  When the
effect of a subhalo can be regarded as a first order perturbation and
the gas is effectively hydrostatic, first order changes in the entropy
distribution produce no first order change in the pressure.  This
makes the perturbations locally isobaric in the presence of thermal
conduction.  Including conduction in the energy equation and expanding
to first order  gives
\begin{equation}
{5 p_0 \over 2 T_0} {\partial \delta T \over \partial t} = \kappa_0
\nabla^2 \delta T,
\end{equation}
where $p_0$, $T_0$, and $\kappa_0$ are the unperturbed pressure,
temperature, and conductivity, respectively, and $\delta T$ is the
temperature perturbation.  From this, we can estimate the time
required for thermal conduction to turn subhalo perturbations from
adiabatic to isothermal on the length scale $\ell$ as
\begin{equation}
t_{\rm cond} = {5 p_0 \ell^2 \over 2 \kappa_0 T_0}.
\end{equation}

As a subhalo moves at speed $v_0$ through the ICM, the gas immediately
around it is continually changing.  The time the subhalo takes to
cross a region of size $\ell$ is $t_{\rm cross} = \ell / v_0$.  For
conduction to make regions of this size isothermal, we must have
$t_{\rm cond} < t_{\rm cross}$.  Equating these timescales determines the
approximate extent, $\ell_{\rm max}$, of the region that is kept isothermal as a subhalo
moves through the ICM,
\begin{equation}
\ell_{\rm max} = {2 \kappa_0 T_0 \over 5 p_0 v_0}.
\end{equation}

Near the center of the Coma Cluster, the electron density is close to
$0.004\rm\ cm^{-3}$ (Lyskova et al., in preparation), so that, for a
temperature of 8 keV, the electron mean free path is $\simeq 3.7$ kpc,
which is small compared to a subhalo.  In the absence of a magnetic
field, the thermal conductivity would be $\kappa_{\rm S} \simeq 4.2
\times 10^{13} \rm\ erg\ cm^{-1}\ s^{-1}\ K^{-1}$.  The effective
conductivity of the magnetized ICM is poorly known, but expressing it
as $\kappa_0 = f \kappa_{\rm S}$, the largest reasonable value is $f
\simeq 0.2$ \citep{nm01}.  Scaling to these values gives 
\begin{eqnarray}
\ell_{\rm max} \simeq 50 && f \left(kT_0 \over 8 \rm\ keV\right)^{5/2}
\left(n_0 \over 0.004\rm\ cm^{-3}\right)^{-1} \nonumber \\ 
&& \times  \left(v_0 \over
1000\rm\ km\ s^{-1}\right)^{-1} \rm\ kpc.
\end{eqnarray}

For $f = 0.2$, this is comfortably smaller than the size of all three
subhalos considered here at their best-fitting distances from the
cluster center (Table~\ref{tab:beta}).  On scales greater than
$\ell_{\rm max}$, the adiabatic approximation is more accurate than
the isothermal approximation.  Because the long range gravitational
impact of a subhalo is most significant for its effects on the surface
brightness, the adiabatic approximation is appropriate for the
subhalos considered here.  Ignoring the effects of conduction slightly
underestimates the peak surface brightness over a subhalo.  We would
need to know both the effective conductivity and the speed of a
subhalo to correct for this.

Note that the emission measure weighted temperature profiles in
Figure~\ref{fig:emwt} are quite sharply peaked and would be more
significantly affected by a high effective thermal conductivity than
the surface brightness profiles.  Potentially, the temperature
profiles can provide a probe of the effective conductivity on
scales of a few tens of kpc, although
existing instruments do not have the sensitivity and accuracy to use
it \citep[cf., e.g.,][]{2003Markevitch}.

\subsection{Unperturbed Cluster Model} \label{sec:beta}

The gas density distribution in the unperturbed cluster is assumed to
have a prolate spheroidal form, with its major axis in the plane of the
sky.  With the $z$ axis along our line of sight, the ellipsoidal
coordinate, $R$, is expressed in terms of cartesian coordinates $(x,
y, z)$, with their origin at the cluster center, by
\begin{displaymath}
\left\{ \begin{array}{l}
x' = x \cos\theta + y \sin\theta;\\
y' = x \sin\theta - y \cos\theta;\\
R^2 = x'^2 + (y'^2 + z^2) / \epsilon^2,\\
\end{array} \right.
\end{displaymath}
where $\theta$ determines the orientation of the major axis and
$\epsilon$ is related to the eccentricity, $e$, by $\epsilon^2 = 1 -
e^2$.  In terms of $R$, the unperturbed gas density distribution is
given by the $\beta$-model for the electron density,
\begin{equation}
n_0 (R) = n_{\rm c}
\left[1+\left(\frac{R}{\rc}\right)^2\right]^{-3\beta /2}, 
\label{density}
\end{equation}
where the central density, $n_{\rm c}$, the core radius, $\rc$, and
$\beta$ are all constants.

The temperature plays a secondary role in our models, only entering
through the unperturbed enthalpy in Equation
(\ref{density_perturbation}).  Thus, the temperature in the unperturbed
cluster is assumed to be constant, an adequate approximation for the
core region of the Coma Cluster \citep[see][]{2001Arnaud,2001Briel}.

\subsection{Subhalo Model} \label{sec:subhalomodel}

The density distribution of gravitating matter in a subhalo is assumed to
have the NFW form \citep{1997NFW}, 
\begin{equation}
\rho(r) = \frac{\rhoscale}{r/\rs(1 + r/\rs)^2},
\label{NFW}
\end{equation}
where $r$ is the distance from the center of the subhalo, $\rs$ is
the scale radius, and $\rhoscale$ is the scale density.  The subhalo
density profile is truncated at $r = \rt$, giving a total mass of
\begin{equation}
\mtot = 4 \pi \rhoscale \rs^3 \left[ \ln (1 + \wt) - {\wt \over 1 + \wt}
  \right],
\label{eqn:mtot}
\end{equation}
where $\wt = \rt/a$.   The corresponding
gravitational potential perturbation is
\begin{equation}
\deltaphi(r) = \cases{\displaystyle
4 \pi G \rhoscale \rs^2 \left[{1 \over 1 + \wt} - {\rs \over r} \ln (1 +
  r/ \rs) \right], & $r \le \rt$\cr
\displaystyle -{G\mtot \over r},& $r > \rt$.}
\end{equation}

The densities of the subhalos determined by \citet{2010Okabe} from
lensing data are higher than expected from simple hierarchical
collapse models.  Using the NFW model, we can estimate the properties
of a progenitor halo from those of its remnant subhalo.  If the subhalo
was simply truncated by falling into the cluster, its mass, $\mtot$,
size, $\rt$, and scale length, $\rs$, determine the scale density,
$\rhoscale$, of its progenitor through Equation (\ref{eqn:mtot}).
In terms of the mean density of the subhalo, this requires
\begin{eqnarray}
\rhot &=& {3 \mtot \over 4 \pi \rt^3} = \rhoscale \left[ {3 \over
    w^3} \left\{\ln (1 + w) - {w \over 1 + w}\right\} \right]_{w = \rt
  / \rs} \nonumber \\
&=& \rhoscale F (\rt / \rs),
\label{eqn:rhot}
\end{eqnarray}
which determines $\rhoscale$ as well as defining the form of $F(w)$.
To estimate the virial radius of the progenitor halo, we adopt the
usual approximation that the mean density within the virial radius is
200 times the critical density, evaluated at the redshift of the Coma
cluster.  That is, the virial radius of the progenitor, $\rvir$, is
determined by solving
\begin{equation}
\rhovir = 200 {3 H^2 \over 8 \pi G} = \rhoscale F (\rvir / \rs) 
= \rhot F (\rvir / \rs) / F (\rt / \rs),
\label{eqn:arv}
\end{equation}
where Equation (\ref{eqn:rhot}) has been used to eliminate
$\rhoscale$.  Here only, $H$ is the Hubble constant at the redshift of
the Coma Cluster (not the specific enthalpy).  The virial mass is then
$4 \pi \rhobar_{\rm vir} \rvir^3 / 3$, an increasing function of
$\rvir$.  For example, for a subhalo with a mass of $10^{13}~\msun$
enclosed within $\rt = 50$ kpc, if its scale radius is $a = 50$ kpc,
the concentration parameter, $c = \rvir / \rs$, of the progenitor
group would have been $\simeq 19.4$ ($\rvir \simeq 970$ kpc) and its
total mass would have been $\simeq 1.07\times10^{14}~\msun$.
Reflecting the high density of the subhalo, a concentration parameter
of 19.4 is higher than expected for such a massive progenitor
\citep[\eg,][]{msh08}.  Since the NFW scale lengths of the subhalos
are largely unconstrained, this issue might be remedied by a different
value for $\rs$.  Differentiating Equation (\ref{eqn:arv}) with
respect to $\rs$ gives the dependence of $\rvir$ on $\rs$ as
\begin{eqnarray}
\left[d \ln F(w) \over d \ln w\right]_{w = \rvir / \rs} &&
  {d \ln \rvir \over d \ln \rs}
= \left[d \ln F(w) \over d \ln w\right]_{w = \rvir / \rs} \nonumber \\
&& - \left[d \ln F(w) \over d \ln w\right]_{w = \rt / \rs}
\end{eqnarray}
and, from the definition of $F$,
\begin{equation}
\left[d \ln F(w) \over d \ln w\right]_{w = r / \rs} = 3 \left[{\rho
    (r) \over \rhobar(r)} - 1\right],
\end{equation}
where $\rhobar(r)$ is the mean density of an NFW halo inside radius
$r$, so that
\begin{equation}
\left[{\rho (\rvir) \over \rhobar (\rvir)} - 1\right] 
  {d \ln \rvir \over d \ln \rs}
= {\rho (\rvir) \over \rhobar (\rvir)} - {\rho (\rt) \over \rhobar
  (\rt)}.
\label{eqn:adep}
\end{equation}

Now $\rho(r) / \rhobar(r) < 1$ and, for an NFW halo, $\rho(r) /
\rhobar(r)$ is a decreasing function of $r$, so that Equation
(\ref{eqn:adep}) shows that $\rvir$ is an increasing function of
$\rs$.  It follows that the virial mass determined from the subhalo
properties is an increasing function of $a$.  Expressing $\rvir$ as
$\rvir = c \rs$, where $c$ is the concentration parameter, Equation
(\ref{eqn:adep}) also shows that $c$ is a decreasing function of
$\rs$.  The progenitor mass cannot be less than the subhalo mass
$\mtot = 10^{13}~\msun$ and a concentration parameter of 19.4 is
already too high for that mass, so reducing $\rs$, which would
increase the concentration parameter, cannot give a consistent
progenitor mass and concentration parameter.  Increasing $\rs$ does
not help either.  Raising $\rs$ to 288 kpc would make the progenitor
mass $10^{15}~\msun$, with a concentration parameter of 7.1, which
is also too high for the halo mass.  Since this progenitor mass is
already about half the total mass of the Coma Cluster
\citep[\eg,][]{ksa07}, it is unreasonable for the progenitor to be any
more massive.  This leaves no reasonable parameters for the progenitor
consistent with expectations for standard $\Lambda$CDM structure
formation models.

So far, we have assumed that the remnant subhalos are simply truncated
by falling into the cluster, but it is far more likely that the mean
density of a subhalo is reduced by its interactions with the cluster.
Allowing for this would require even greater mean densities for the
progenitors, exacerbating the issue of the high subhalo densities.
Thus, the subhalos defined by \citet{2010Okabe} are smaller than
expected from standard structure formation models, if they fell into
the Coma Cluster in the recent past.  The simplest resolution to this
issue would be if the more massive central subhalos fell into Coma
long in the past, when the mean densities of their progenitors would
have been higher \citep[cf.][]{1996Biviano}.  Our results are
much less sensitive to the structure of a subhalo than they are to its
total mass, so we do not pursue this issue any further here.

\subsection{Estimate for the Effect of a Subhalo} \label{sec:estimate}

\begin{figure*}[hbt!]
\centerline{%
\raisebox{0.05\textwidth}{\includegraphics[width=0.4\textwidth]{%
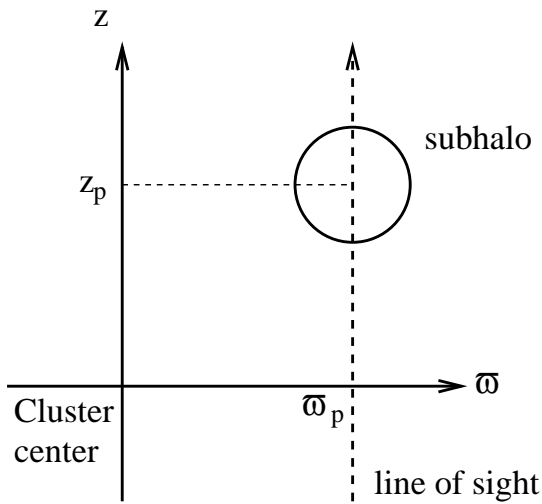}}%
\hbox to 0.05\textwidth{ }%
\includegraphics[width=0.47\textwidth]{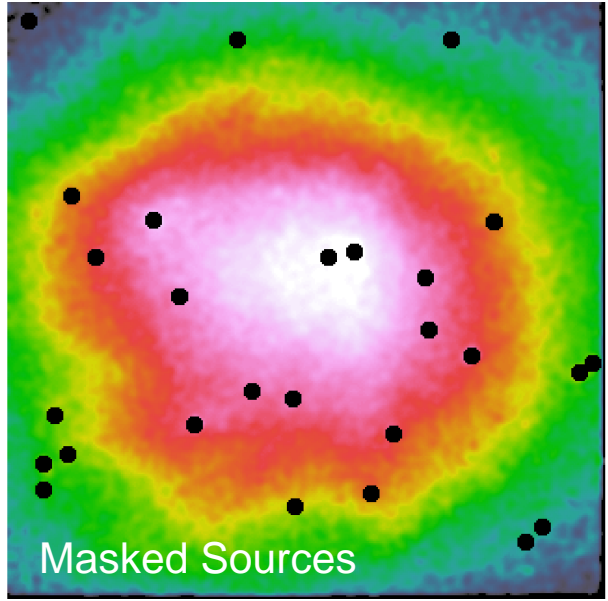}}
\caption{Left: parameters defining the location of a
  subhalo.  Right:  
XMM-Newton image showing the mask used to remove the point sources.}
\label{mask}
\label{fig:subhalo}
\end{figure*}

With the approximations outlined above, the perturbation to the X-ray
surface brightness must still be computed numerically (Section
\ref{sec:Num_approach}). However, before considering the numerical results, we
make a crude estimate of the effect on the X-ray surface brightness
of a single subhalo.  Under our assumptions, the fractional change in
X-ray surface brightness is
\begin{equation}
{\delta I \over I} = {\int (n^2 - n_0^2) dz \over \int n_0^2 dz},
\label{eqn:fracsb}
\end{equation}
where the integrals are along the line of sight.  To first order in
$\deltaphi$, Equation (\ref{density_perturbation}) gives $n^2
- n_0^2 \simeq n_0^2 (-3 \deltaphi / H_0)$.  For a localized perturbation,
we get the further approximation,
\begin{eqnarray}
&&\int (n^2 - n_0^2) dz \simeq 3 n_0^2 / H_0 \int_{-\zmax}^{\zmax}
-\deltaphi dz \nonumber \\ 
&&\simeq 6 n_0^2 {G \mtot \over H_0} \ln \left[{\zmax \over \varpi} +
  \sqrt{1 + \left(\zmax \over \varpi \right)^2} \right],
\end{eqnarray}
where $\varpi$ is the minimum distance of the line of sight from the
center of the subhalo and this result applies to lines of sight with
$\varpi > \rt$.  The integration has been truncated at $z = \pm\zmax$
to avoid a logarithmic divergence (a spurious consequence of the crude
approximations used here).  For lines of sight that intersect the
subhalo, \ie, for $\varpi \le \rt$, the magnitude of the result is
similar.  Setting the logarithmic factor to unity and using the form
(\ref{density}) for the unperturbed ICM density, $n_0$, finally gives
us an order of magnitude estimate for the fractional perturbation in
the surface brightness near the center of the subhalo
\begin{equation}
{\delta I \over I} 
\simeq {6 GM_t \over \rc H_0 B(3\beta - 1/2, 1/2)} \times
{\rc (\rc^2 + \rp^2 )^{3\beta - 1/2}  \over (\rc^2 + \rp^2 + \zp^2)^{3\beta}},
\label{enhancement_text}
\end{equation}
where $B(a,b)$ is the $\beta$ function, and $\rp$ and $\zp$ are
cylindrical coordinates for the center of the subhalo measured with
respect to the cluster center, \ie, their projected separation and
their separation along our line of sight, respectively
(Figure~\ref{fig:subhalo}).  
For an unperturbed gas temperature of $\kb T_0 = 8$ keV, a subhalo
mass of $M_t = 10^{13}~\msun$, and $\beta$-model core radius of $\rc =
300$ kpc, the first factor on the right in Equation
(\ref{enhancement_text}) is $\simeq 0.17$ for $\beta \simeq 2/3$.
Thus, such a subhalo should be detectable in X-ray observations,
unless it is far from the
cluster center ($|\zp|$ or $\rp \gg \rc$).  Note that the estimate for
$\delta I/I$ in Equation (\ref{enhancement_text}) only depends on the
mass of a subhalo, not on its size or scale parameters.  This reflects
the insensitivity of the full, numerical results to details of the
mass distribution in the subhalos.  It is a consequence of the long
range of the gravitational force.

\section{Simulated X-ray Images of the Coma Cluster} \label{sec:Num_approach}

In this section we present numerical results, employing the model
outlined above to compute the effects of subhalos on the X-ray surface
brightness of the Coma Cluster.  The X-ray surface brightness is
computed by integrating the square of the gas density, $n$, given by
Equation (\ref{density_perturbation}), along lines of sight through a
model cluster with an unperturbed gas temperature of $\kb T_0 = 8$ keV
and an unperturbed gas density profile given by the $\beta$-model
(Section \ref{sec:beta}).  Most parameters of the $\beta$-model are
determined in the fits, as outlined below.  Gas densities are
evaluated and integrated numerically on a rectangular parallelepiped
of $321 \times 321 \times 1000$ cubic cells, 1.906 kpc on
a side and centered on the cluster.  The long axis of the grid is
parallel to the line of sight.

Subhalos are modeled with NFW profiles (Section
\ref{sec:subhalomodel}), using parameters from \citet{2010Okabe}.  In
the following analysis, apart from Section~\ref{sec:struct}, the NFW
scale radius, $\rs$, is set to 50 kpc in a subhalo that is also
truncated at $\rt = 50$ kpc to be consistent with the lensing results.
The magnitude of the change in surface brightness due to a subhalo
increases with the density of the surrounding gas, so that it is
maximized when the subhalo is closest to the cluster midplane, defined
by $\zp = 0$.  Because the effect of a subhalo is sensitive to its
distance from the midplane, we can constrain the locations 
of the subhalos along our line of sight.

\subsection{Fitting} \label{methods}

Models are multiplied by the exposure map and added to the background
image for fitting to the raw 0.5--2.5 keV \xmm{} image of the
central part of the Coma Cluster, made from the combined EPIC MOS
data.  Goodness of fit is determined using $\chi^2$ as the fit
statistic.  The image was binned into $2''\times 2''$ pixels and the region
fitted had an extent of $10'\!.7 \times 10'\!.7$.  The error estimate
for each pixel includes Poisson errors for the raw image and for the
background data.  The fit was masked to exclude point sources evident
in the \xmm{} image (Figure~\ref{mask}).  Away from point sources, the
average photon count per pixel in the raw image varies from over 50
near the center to about 8 at the periphery.

Models include the three subhalos marked in Figure~\ref{clumps_okabe}.
The centers of the subhalos corresponding to the giant elliptical (gE)
galaxies NGC~4889 and NGC~4874 were placed on the sky at their
respective optical centers, rather than at the locations determined
from the lensing data of \citet{2010Okabe}. The center of
  subhalo 3 was placed at the location of the X-ray bright source 
closest to the lensing position, since the X-ray surface brightness 
is elongated toward this position.  The lensing positions are
consistent with these positions, within the errors of the lensing
measurements, but using the latter positions gives somewhat better
fits ($\Delta \chi^2 \simeq 700$, Table~\ref{tab:beta}).  
  Coordinates used for the subhalo
centers are, for NGC~4889 (J2000) =
$(195.034^\circ, 27.977^\circ)$, for NGC 4874, $(194.899^\circ,
27.959^\circ)$, and for the southern mass concentration,
$(195.006^\circ, 27.855^\circ)$.  Fixed parameters for the subhalo
models are given in Table~\ref{tab:subhalos}.  The location of the
center of the X-ray emission from the Coma Cluster was determined by
fitting a circularly symmetric $\beta$-model to the central region of the
cluster.  This center, $(194.932^\circ, 27.929^\circ)$, was kept fixed
in all subsequent fits.  All other parameters for the cluster
$\beta$-model ($\theta$, $\epsilon$, $\rc$, $\beta$, normalization) are
determined in the fits.  For each subhalo, the only free parameter is
its location along our line of sight, measured by its distance from
the cluster midplane ($\zp$ in Figure~\ref{fig:subhalo}).

\begin{deluxetable}{cccccc}
\tablecaption{Parameters of Subhalo Models} 
\tablewidth{0pt} 
\tablehead{ 
\colhead{Subhalo} &
\colhead{$\alpha_{\rm J2000}$} &
\colhead{$\delta_{\rm J2000}$} &
\colhead{$\mtot$} &
\colhead{$\rs$} &
\colhead{$\rt$} \\
\colhead{} &
\colhead{(deg)} &
\colhead{(deg)} &
\colhead{($10^{12}~\msun$)} &
\colhead{(kpc)} &
\colhead{(kpc)}
}
\startdata
1, NGC~4889 & 195.034 & 27.977 & 11.0 & 50 & 50 \\
2, NGC~4874 & 194.899 & 27.959 & 6.57 & 50 & 50 \\
3 & 195.006 & 27.855 & 7.57 & 50 & 50 \\[-6pt]
 \enddata
\label{tab:subhalos}
\end{deluxetable}

Fitting models with up to eight nonlinear, free parameters with
$\sim10^5$ degrees of freedom (dof) presented some challenges.  To ensure
that the global best fit was found, two different fitting procedures,
a Levenberg-Marquardt algorithm \citep{br03} and a grid search, were
employed to cross check one another.  Initially, the three subhalos
were placed at $z = 1$ Mpc, effectively removing them from the fit,
and the best fitting parameters for the $\beta$-model alone were
determined.  This fit gives $\beta = 0.80$, $\rc = 351$ kpc,
$\epsilon = 0.83$, $\theta = -3.9^\circ$, and $\chi^2 = 139204.6$,
or a reduced $\chi^2$ ($\chi^2$/dof) of 1.3881 (Table~\ref{tab:beta}).

\begin{figure*}[ht!]
\begin{center}
\includegraphics[width=1.00\textwidth]{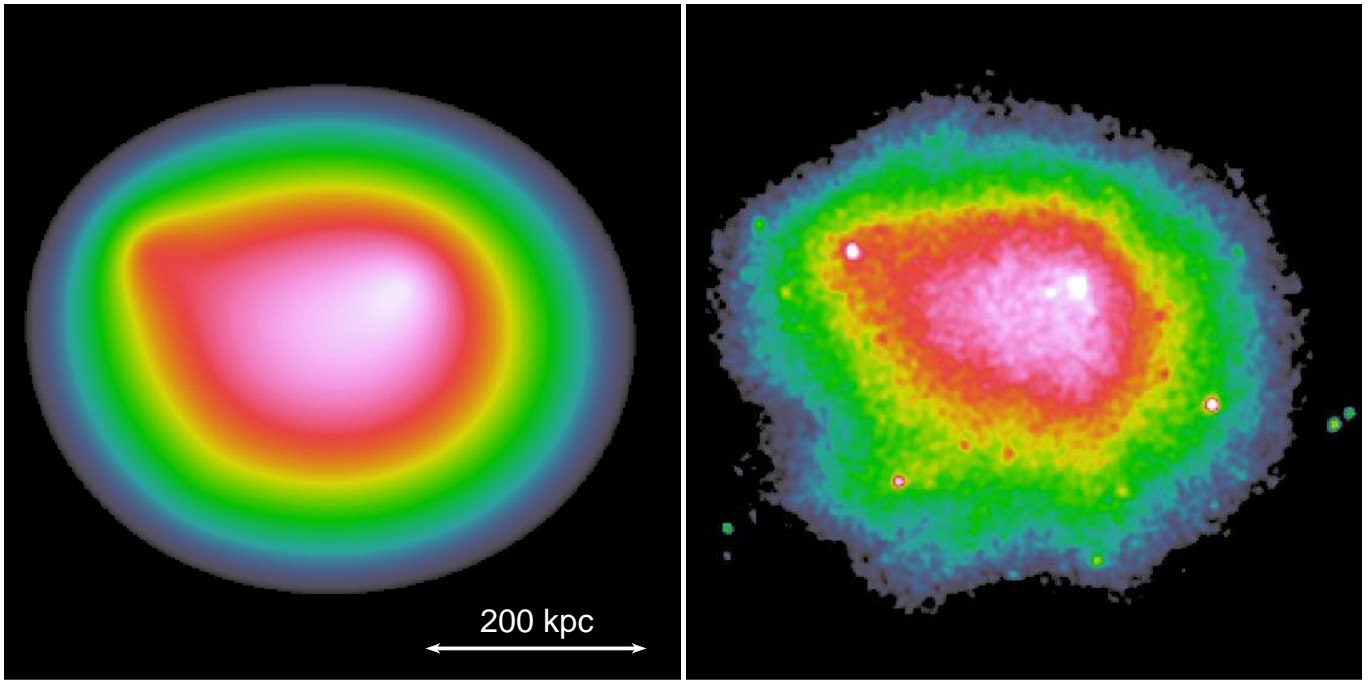}
\end{center}
\caption[Model vs. Real Image]{Left: Model X-ray image for
  the Coma Cluster, made 
  by embedding three dark matter subhalos in the elliptical $\beta$-model
  of Section \ref{sec:beta}. Right: 0.5--4 keV \xmm{}
  X-ray image of the Coma Cluster.  North is up and east to the left.
  Increased brightness due to the presence of the
  subhalos elongates the model image in the direction of each subhalo,
  particularly toward the east.  These features correspond
  well to structure in the X-ray image.
}\label{model_vs_real}
\end{figure*}

When the three subhalos are included in the fit, with subhalos 1 and 2
at the locations of the two gE galaxies, allowing the
positions along the line of sight of the three subhalos, $z_i$ for $i
= 1, 2, 3$, to vary simultaneously gives the best fitting parameters
$z_1 = 121$ kpc, $z_2 = 7.9$ kpc, $z_3 = 399$ kpc, $\rc = 387$ kpc,
$\theta = -0.1^\circ$, $\epsilon = 0.84$, for which $\chi^2 =
119114.7$, making the reduced chi squared $\chi^2/\dof = 1.1878$.  The
improvement in the fit, of $\Delta\chi^2 = 20089.9$ for the addition
of three parameters, is highly significant.  Visual comparison of the
model and cluster images in Figure~\ref{model_vs_real} shows that the
subhalos, particularly the two associated with NGC~4889 and NGC~4874,
make the model X-ray image very similar to the observed cluster.
Using $\Delta\chi^2 = 2.706$ to estimate confidence ranges for the
three subhalo positions, with all of the other fitted parameters free,
gives 90\% confidence ranges  of $z_1 = 121 \pm 5$ kpc, $z_2 =
7.9 \pm 7.5$ kpc, and $z_3 = 399 \pm 10$ kpc.

\begin{deluxetable*}{cccccccccc}
\tablecaption{Parameters of X-Ray Surface
  Brightness Model} \tablewidth{0pt} \tablehead{ \colhead{Model$^{\it a}$} &
  \colhead{$\beta$} & \colhead{$\epsilon$} & \colhead{$\theta$$^{\it b}$} &
  \colhead{$\rc$} & \colhead{$z_1$} & \colhead{$z_2$} &
  \colhead{$z_3$} &
\colhead{dof} &
\colhead{$\chi^2$} \\
\colhead{} &
\colhead{} &
\colhead{} &
\colhead{(deg)} &
\colhead{(kpc)} &
\colhead{(kpc)} &
\colhead{(kpc)} &
\colhead{(kpc)} &
\colhead{} &
\colhead{}
}
\startdata
Cluster only$^{\it c}$ & 0.80
&0.83 & -3.9 & 351 &1000 & 1000 & 1000 
& 100283 & 139204.6 \\
Subhalos (lensing)$^{\it d}$ & 0.82 
&0.84 & 3.2 & 395 & 39 & 5.8 & 361
& 100280 & 119817.0 \\ 
Subhalos (gE)$^{\it e}$ & 0.80 
&0.84 & -0.1 & 387 & 121 & 7.9 & 399
& 100280 & 119114.7 \\[-6pt]
\enddata
\tablecomments{$^{\it a}$Using the cluster $\beta$-model of Section
  \ref{sec:beta} and subhalo models of Section \ref{sec:subhalomodel}.
  $^{\it b}$Position angle of the major axis, increasing north of west.
  $^{\it c}$Subhalos fixed at 1 Mpc from the cluster midplane.
  $^{\it d}$Subhalos positions on our line of sight free to vary, with the
  subhalos centered at the lensing positions.
  $^{\it e}$As for $d$, but with subhalos 1 and 2 centered at the
  positions of the two gE's and subhalo 3 centered on the X-ray bright
  source.
 }
\label{tab:beta}
\end{deluxetable*}

\subsection{Structure and Masses of the Subhalos} \label{sec:struct}

\begin{figure}[hbt!]
\begin{center}
\includegraphics[width=0.475\textwidth,bbllx=24,bblly=162,bburx=564,bbury=586]{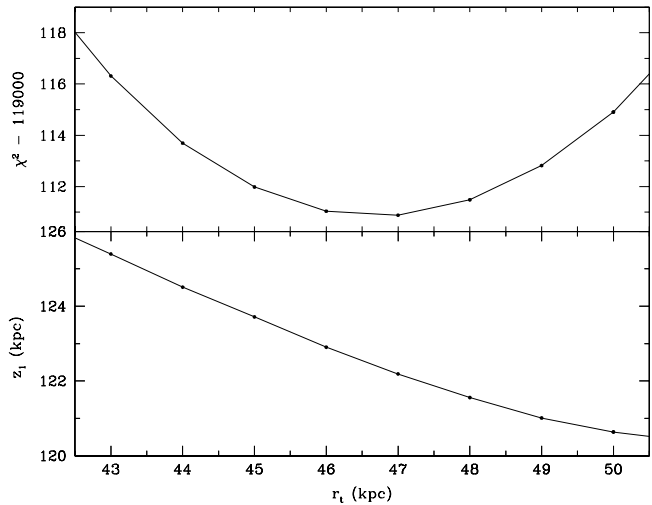}
\end{center}
\caption{Effect of varying the truncation radius of subhalo 1.  This
  plot shows how the fit depends on the structure of subhalo 1,
  through its truncation radius, $\rt$.  The lower panel shows the
  best fitting distance of the subhalo from the cluster midplane along
  our line of sight, while the upper panel shows $\chi^2 - 119000$.
  For these fits, the scale radius of the subhalo, $\rs$, is tied to
  its truncation radius, $\rt$, the subhalo mass is fixed and the
  parameters of the other subhalos are frozen.  The remaining fit
  parameters are free.}
\label{fig:rt1}
\end{figure}

As discussed in Section \ref{sec:subhalomodel}, the subhalo properties
determined from the lensing measurements give them higher densities
than expected for hierarchical collapse models.  To test how sensitive
our results are to the structure of the subhalos, we varied the
truncation radius of the most massive subhalo, corresponding to
NGC~4889 (subhalo 1),  while its mass was kept fixed.  Since the
NFW model is unlikely to be valid for a truncation radius, $\rt$, smaller
than the NFW scale length, $\rs$, we tied these two parameters
together for these fits ($\rs = \rt$). 
All parameters for the other subhalos were
frozen at their best-fitting values, since the fits are not sensitive
to these.  The resulting $\chi^2$ and positions along our line of
sight for subhalo 1 are plotted against $\rt$ in Figure~\ref{fig:rt1}.
The fit is clearly affected by the structure of this subhalo, with a
moderately significant improvement of $\Delta\chi^2 \simeq 4$ for the
best fit.  However, at 47 kpc, the best fitting truncation radius for
subhalo 1 is actually smaller than the value of 50 kpc from the
lensing results, only exacerbating the problems caused by the high
density of this subhalo (Section \ref{sec:subhalomodel}).  This result
depends on our model through properties that are not well determined,
so it should be treated with caution.  For example,
Table~\ref{tab:beta} shows that the best fitting position, $z_1$, for
the NGC~4889 subhalo is quite sensitive to the position on the sky of
this subhalo.  Nevertheless, the fact that the subhalo structure has
an appreciable effect on the fit shows that the X-ray data could
potentially be used to probe its structure.  Results for the NGC~4874
subhalo are similar.

From Section~\ref{sec:estimate}, the X-ray surface brightness of the
model cluster is expected to be considerably more sensitive to the
total mass of a subhalo than to its structure.  In principle, we can
determine the subhalo masses by fitting the X-ray image, using data
that is completely independent of the lensing mass measurements.  To
investigate the agreement between the lensing and X-ray mass
determinations, we have again used subhalo models with $\rs = \rt$
allowed to vary, but here the scale density, $\rho_0$, or,
equivalently, the mean density of the subhalo, is kept fixed, so that
its mass varies as $\propto \rt^3$.  Parameters for the cluster model
are free to vary, but those for the other subhalos are kept fixed at
their best-fitting values (last row of Table~\ref{tab:beta}). 

\begin{figure*}[hbt!]
\centerline{%
\includegraphics[width=0.47\textwidth,bbllx=22,bblly=164,bburx=564,bbury=584]{%
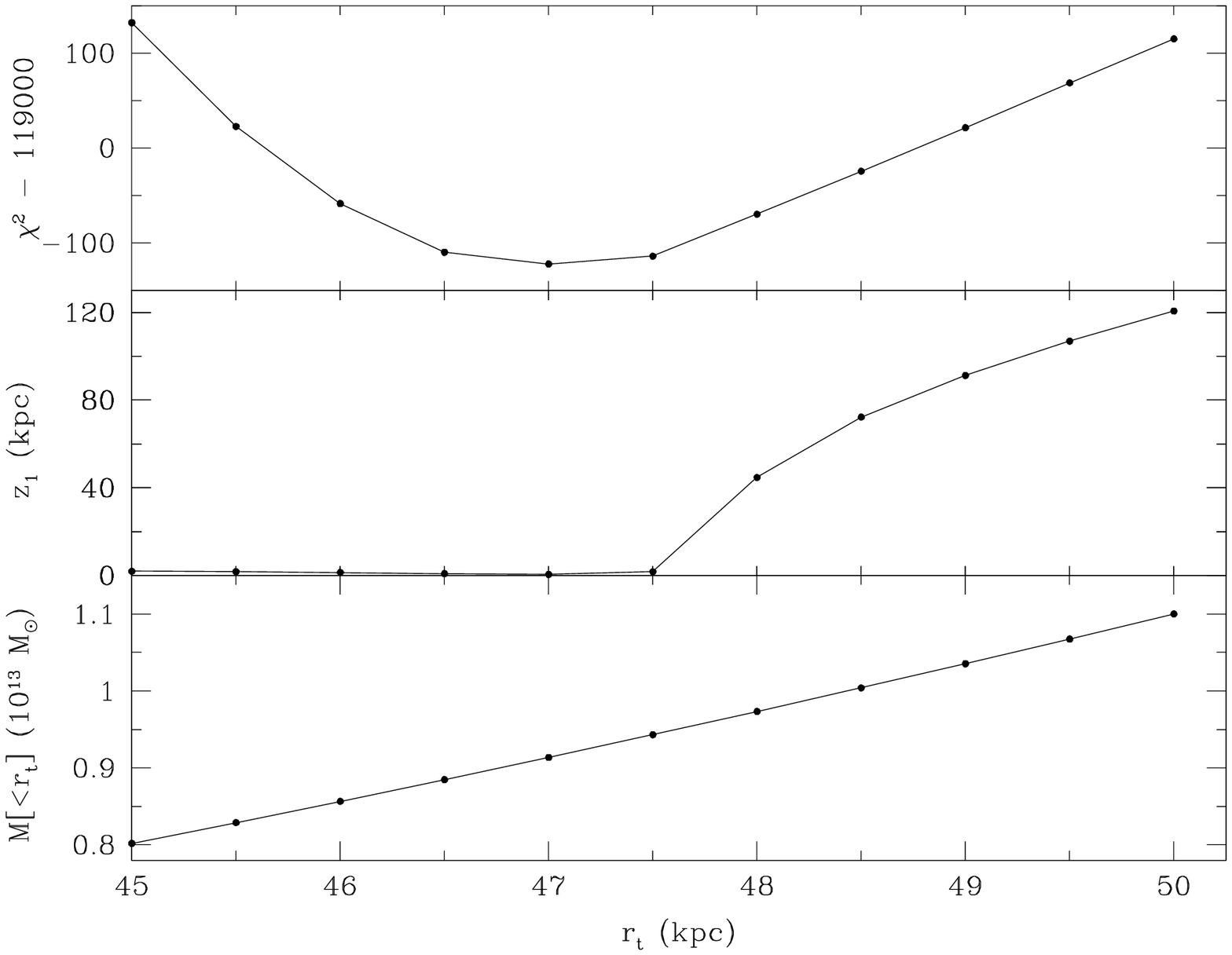}
\includegraphics[width=0.47\textwidth,bbllx=22,bblly=164,bburx=564,bbury=584]{%
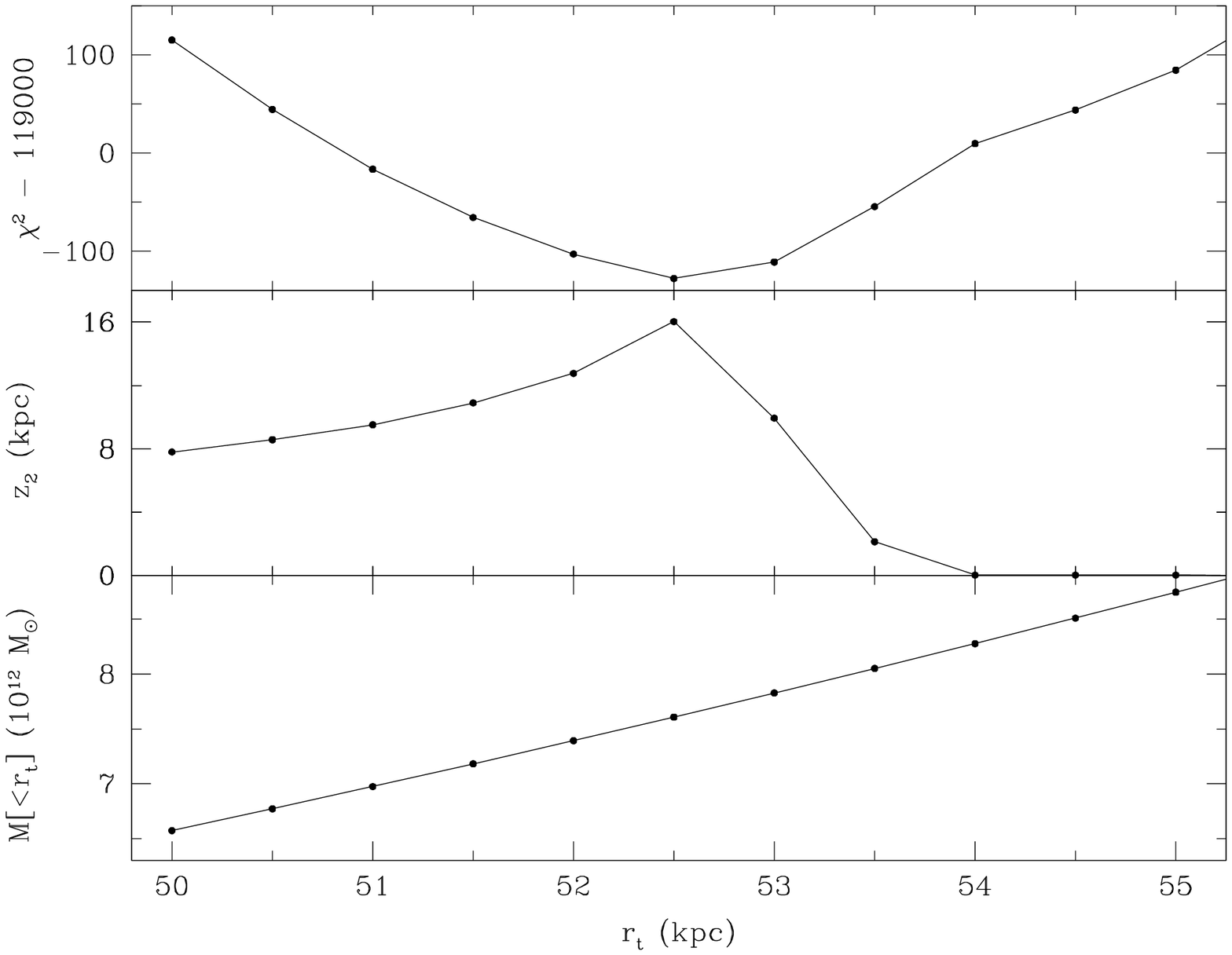}}
\caption{Effect on the fit of varying subhalo mass.  Left: Fit
  results for the NGC~4889 subhalo are plotted against its truncation
  radius, $\rt$, with the mean density of the subhalo held constant.
  Right: same for NGC~4874.  The bottom panels show the subhalo
  mass ($\propto \rt^3$), the middle panels show its position along our
  line of sight, and the top panels show $\chi^2 - 119000$.  Model
  parameters match the lensing parameters in Table~\ref{tab:subhalos}
  for $\rt = 50$ kpc.  For these fits, the scale radius of the
  subhalo, $\rs$, is tied to its truncation radius, $\rt$, the mean
  density of the subhalo is fixed, while the parameters of the other
  subhalos are frozen.}
\label{fig:mass}
\end{figure*}

The left panel of Figure~\ref{fig:mass} shows results for varying the
mass of the NGC~4889 subhalo (subhalo 1).  Plotted against $\rt$,
the subhalo mass is given in the bottom panel, distance from the cluster
midplane in the middle panel, and $\chi^2$ in the top panel.  Reducing
the mass of the subhalo decreases its impact on the gas, hence the
X-ray surface brightness, but this can be offset by shifting the
subhalo toward the Coma Cluster midplane, where the unperturbed gas
density is greater.  The best-fitting mass for subhalo 1 is $9.14
\times 10^{12}~\msun$,  $\simeq 1.2\sigma$ lower than the lensing mass of
\citet{2010Okabe}.   The corresponding improvement in the fit
is $\Delta\chi^2 \simeq 237$.  Results for varying the mass of the
NGC~4874 subhalo are shown in the right panel of
Figure~\ref{fig:mass}.  Here, the best fitting mass of
$7.61\times10^{12}~\msun$ is $\simeq 0.9 \sigma$ greater than
the lensing mass and the improvement in the fit is $\Delta\chi^2
\simeq 242$ (see Table \ref{tab:subhalos_masses}).  This subhalo starts from much closer to the cluster
midplane than subhalo 1 and the nonmonotonic variation of $z_2$ seen
in the plot results from the interplay between the subhalo and cluster
parameters ($\theta$ and $\epsilon$ in particular, see Section
\ref{sec:beta}).  The mass of subhalo 3 also affects the fit
significantly, but $\chi^2$ continues to improve as it is increased
well beyond the range consistent with the lensing data (\eg, doubling
the subhalo mass moved it to $z_3 \simeq 510$ kpc from the midplane
and an improvement of $\Delta\chi^2 \simeq 350$).  This indicates that
our model assumptions are poorer for this subhalo.

\begin{deluxetable*}{ccccccc}
\tablecaption{Best-fit Masses of the Subhalos} 
\tablewidth{0pt} 
\tablehead{ 
\colhead{Subhalo} &
\colhead{$\alpha_{\rm J2000}$} &
\colhead{$\delta_{\rm J2000}$} &
\colhead{$\mtot$} &
\colhead{$\rs$} &
\colhead{$\rt$} &
\colhead{$\Delta\chi^2$} \\
\colhead{} &
\colhead{(deg)} &
\colhead{(deg)} &
\colhead{($10^{12}~\msun$)} &
\colhead{(kpc)} &
\colhead{(kpc)} &
\colhead{}
}
\startdata
1, NGC~4889 & 195.034 & 27.977 & 9.1 & 47 & 47 & 237 \\
2, NGC~4874 & 194.899 & 27.959 & 7.6 & 52.5 & 52.5 & 242 \\[-6pt]
 \enddata
\label{tab:subhalos_masses}
\end{deluxetable*}

\subsection{Temperature Perturbations} \label{sec:temps}

\begin{figure}[hbt!]
\begin{center}
\includegraphics[width=0.475\textwidth,bbllx=24,bblly=162,bburx=564,bbury=502]{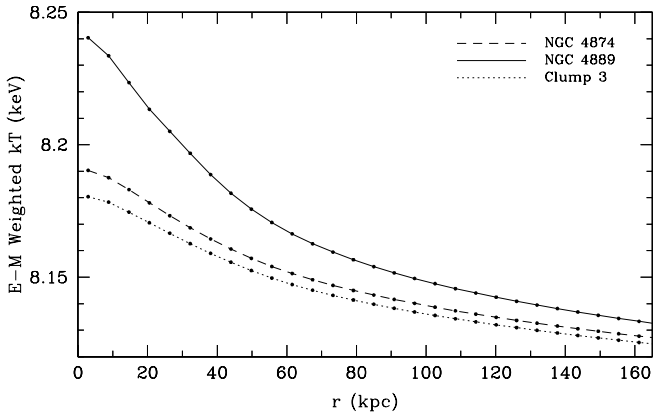}
\end{center}
\caption{Emission measure weighted temperature profiles computed
  from our model for the three subhalos.  
  The peak temperature over the NGC~4889 subhalo
  is 3\% greater than the model temperature of 8 keV.} 
\label{fig:emwt}
\end{figure}

Next to the gravitational effect of subhalos, the most likely cause of
large-scale enhancements in the X-ray surface brightness of the Coma
Cluster is low entropy gas that has fallen in during the ongoing
collapse.  In that case, the pressure of the gas responsible for the
excess emission would be similar to that of its surroundings, but its
density would be higher, boosting the X-ray brightness.  By contrast,
gas compressed adiabatically in the potential of a subhalo should be
hotter than its surroundings.  For the model used here, the
temperature enhancement due to adiabatic compression can be determined
from $H = H_0 - \deltaphi$ (Equation (\ref{eqn:pert})).  Using the
emission measure weighted temperature to estimate the fractional
temperature perturbation, the model gives
\begin{equation}
{\delta T \over T_0} = {\int n^2 H \, dz \over H_0 \int n^2 \, dz} - 1
\simeq {\int n_0^2 (- \deltaphi/H_0) dz \over \int n_0^2 dz}
\simeq {1 \over 3}{\delta I \over I},
\end{equation}
where the third form is obtained by expanding to first order in
$\deltaphi/H_0$, using Equation (\ref{density_perturbation}), and the
last form is obtained in the same manner from the expression for the
surface brightness fluctuations in Equation (\ref{eqn:fracsb}).
Combining this result with the rough estimate for $\delta I/I$ from
Section~\ref{sec:estimate}, the maximum temperature fluctuation due to
a subhalo is expected to be roughly 5\%.  Evaluated numerically for
the best-fitting parameters in the last line of Table~\ref{tab:beta},
the maximum temperature rise is only about 3\% for NGC~4889 and less
for the other two subhalos (Figure \ref{fig:emwt}).  These are not
large enough to detect in small regions with the current data,
particularly at the relatively high temperature of the Coma Cluster.

\begin{figure*}[hbt!]
\begin{center}
\includegraphics[width=\textwidth]{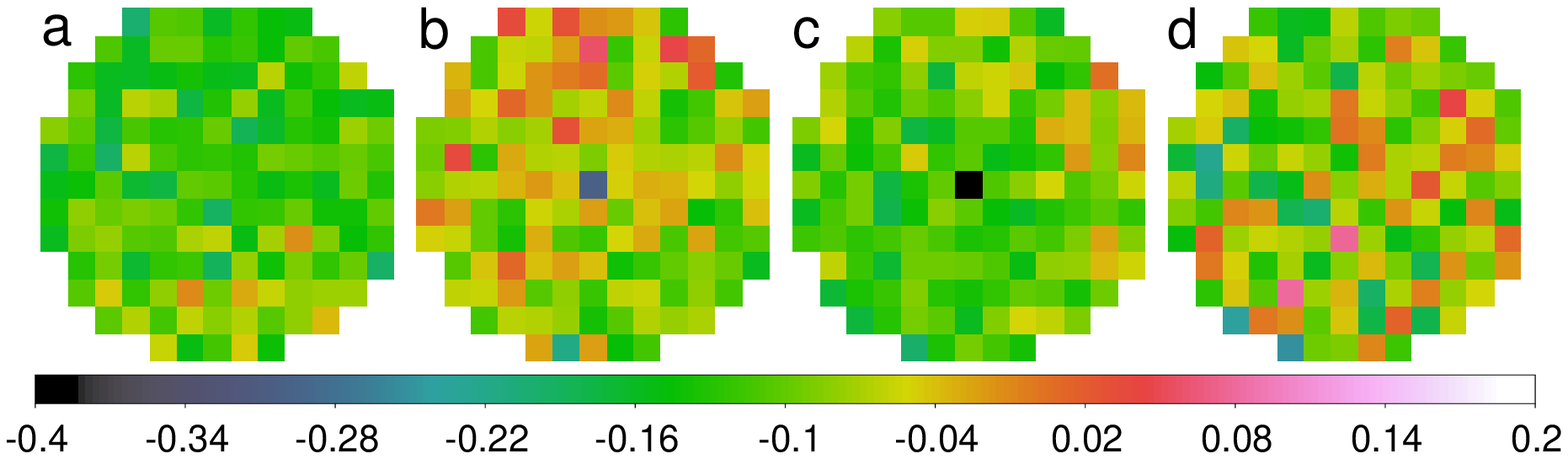}
\end{center}
\caption{Hardness-ratio maps for the subhalos.  \textit{a)} a
  representative region lying between the 3 subhalos.  \textit{b)}
  subhalo 1, centered on NGC~4889. \textit{c)} subhalo 2, centered on
  NGC~4874.  \textit{d)} subhalo 3, centered on the X-ray bright
  source.
  The hardness ratio is defined in the text.  Each map covers a region
  $102''$ or $\simeq 50$ kpc in radius.  Apart from the single cool
  central pixels for NGC~4889 and NGC~4874, there is no significant sign of a
  radial temperature gradient in any of these regions.}
\label{HR_temps}
\end{figure*}

\begin{figure*}[hbt!]
\centerline{%
\includegraphics[width=0.47\textwidth,bbllx=21,bblly=162,bburx=564,bbury=498]{%
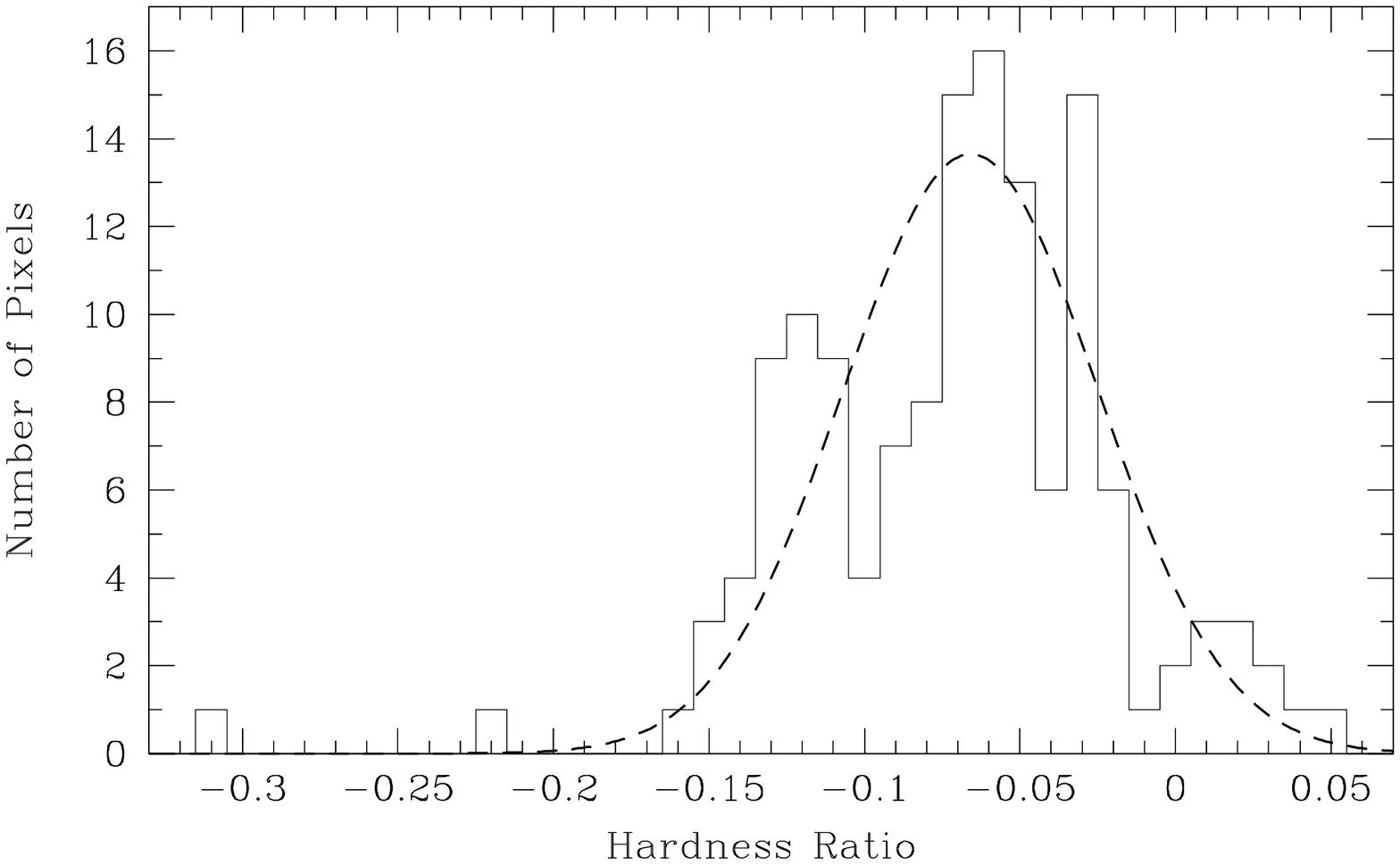}
\includegraphics[width=0.47\textwidth,bbllx=21,bblly=162,bburx=564,bbury=498]{%
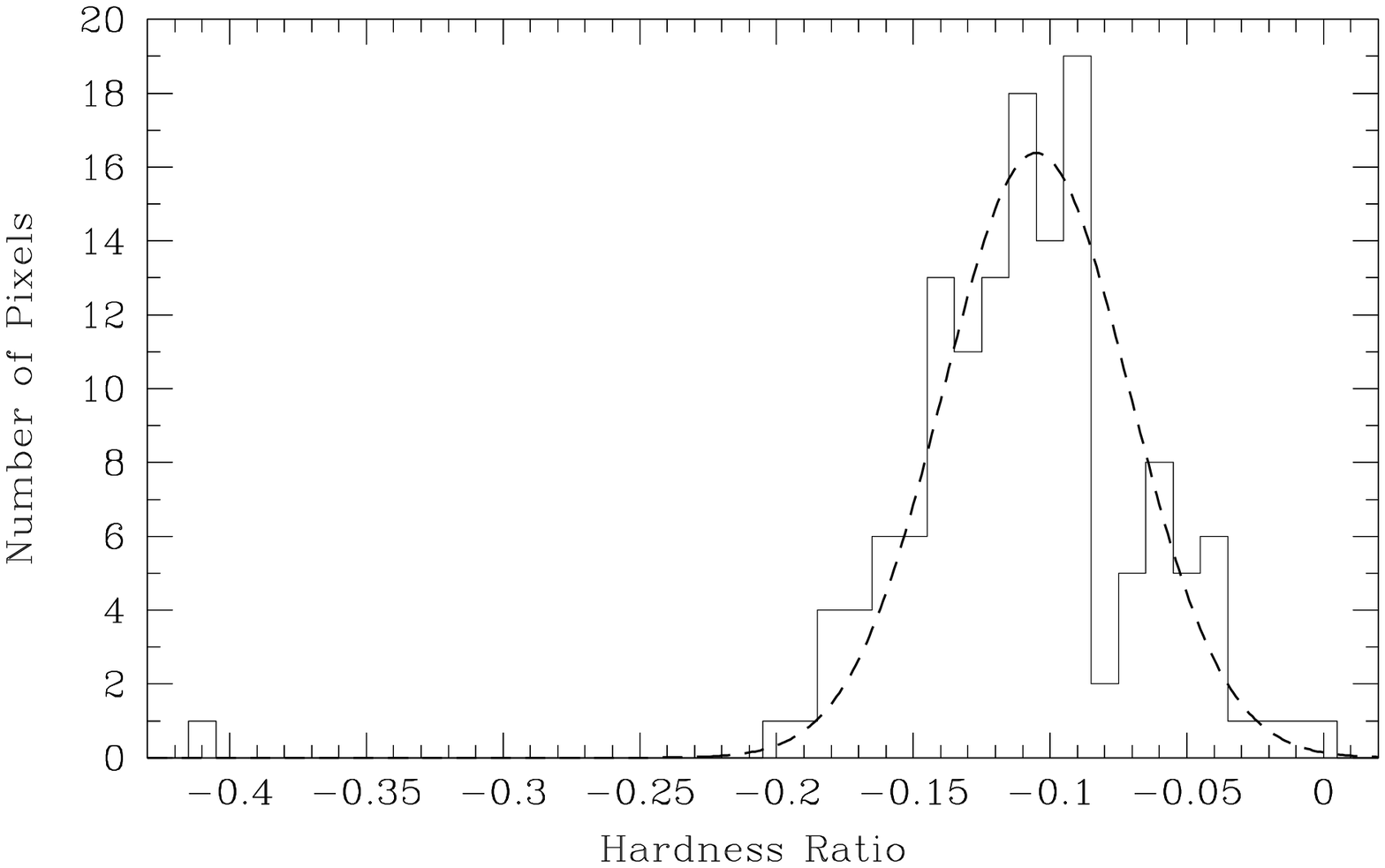}}
\caption{Histograms of the hardness ratio for regions b (NGC~4889,
  left panel) and c (NGC~4874, right panel) of Figure~\ref{HR_temps}.
  The dashed lines show Gaussians with the same means as the data, but
  with standard deviations calculated to include only the effect of
  Poisson noise on the hardness ratio.  Apart from a single cool
  central pixel at the left in each histogram, they are reasonably
  consistent with a constant hardness ratio in each region.}
\label{fig:hrhist}
\end{figure*}

However, we can check for signatures of cooler gas that might be
responsible for the enhanced X-ray emission.  For this purpose, we
have made maps of the hardness ratio in 50 kpc regions around each of
the subhalo locations by binning \chandra{} data for the bands
0.3--1.5 keV and 1.5--6.0 keV into $15.\!''74 \simeq 7.5$ kpc pixels.
The images in Figure~\ref{HR_temps} show $\rm(hard - soft)/(hard +
soft)$.  Apart from the innermost pixels for NGC~4889 and NGC~4874,
there is no indication of cooler gas in any of the subhalos.
\citet{2001Vik} showed that there are $\simeq 3$ kpc cores of 1--2
keV gas in both of these gE galaxies, which affect the
central pixel of the respective hardness-ratio images.  These regions
are masked out of the image used to fit the models and they certainly
do not account for the extended regions of enhanced X-ray emission
associated with the subhalos.  There are no other signs of cooler gas
in the images of Figure \ref{HR_temps}, consistent with previous
investigations \citep{2001Arnaud, 2001Briel, 2003Neumann}.  In
particular, there is no sign of a radial temperature increase in any
of these regions.  Apart from the solitary low pixels for the two cool
cores, the histograms of the hardness ratio for each of these regions
are reasonably consistent with a constant value.  This is illustrated in
Figure~\ref{fig:hrhist}, which shows histograms of the hardness ratio
for the regions around NGC~4889 (left panel and panel b) and NGC~4874
(right panel and panel c) in Figure~\ref{HR_temps}.  Statistical noise
in the soft and hard counts contributes to the standard deviation of
the hardness ratio as
\begin{equation}
\sigma_{\rm HR, stat} \simeq {2 \sqrt{h^2 \sigma_s^2 + s^2 \sigma_h^2}
  \over (h + s)^2},
\label{equation:HRstat}
\end{equation}
where $s$ and $h$ are the mean counts for the soft and hard bands,
respectively, and $\sigma_s$ and $\sigma_h$ are the respective
standard deviations for the two bands.  Dashed lines in
Figure~\ref{fig:hrhist} show Gaussians with means, $\mu_{\rm HR}$,
equal to the data mean and normalized
to match the number of samples with standard deviations 
given by Equation (\ref{equation:HRstat}). The plotted number of samples per
bin is given by
\begin{equation}
N_{{\rm samples}}={N w \over \sigma_{\rm HR, stat} \sqrt{2 \pi}} \exp \left(- {(\psi -
  \mu_{\rm HR})^2 \over 2 \sigma_{\rm HR, stat}^2} \right),
\end{equation}
where $\psi$ is the hardness ratio, $N$ is the total number of samples, and
$w$ is the width of the histogram bins.  The agreement between these
Gaussians and the histograms shows that, apart from the single cool
pixel in each core, the histograms show little evidence for variation
of the hardness ratio over each subhalo.


\section{Discussion} \label{sec:discuss}

Under our model assumptions, the distances from the cluster midplane
of each of the three central subhalos are well constrained at the 90\%
confidence level ($z_1 = 121 \pm 5$ kpc, $z_2 =
7.9 \pm 7.5$ kpc and $z_3 = 399 \pm 10$ kpc),
providing useful information on their three-dimensional
locations within the Coma Cluster (Section \ref{methods}).  In
particular, subhalo 1 (NGC~4889) is closer to the cluster midplane
than its distance from the cluster center and subhalo 2 (NGC~4874) is
even closer to the cluster midplane.  By contrast, subhalo 3 is 1.7
times further from the midplane than it is from the cluster center.
These results are affected by our model assumptions, particularly
those concerning the masses and locations on the sky of the
subhalos.  They are also affected by our largely arbitrary assumption
that the gas distribution in the unperturbed cluster is prolate, with
its major axis in the plane of the sky.  While these assumptions
affect the distances of the subhalos from the cluster midplane, they
do not alter the qualitative result that subhalos 1 and 2 lie close to
the cluster midplane, while subhalo 3 does not.  Our results are
consistent with previous findings that NGC~4889 and NGC~4874 lie close
to the cluster center \citep[e.g.][]{1996Biviano}.

Nonlinearity, from Equations (\ref{density_perturbation}) and
(\ref{eqn:fracsb}), means that the impact of two or more subhalos on
the X-ray surface brightness increases when they are closer together.
As a result, the placement of subhalos about the cluster midplane can
alter the fit.  This is seen in the fit results for subhalos 1 and 2.
In computing the confidence range for subhalo 2, we found that
$\Delta\chi^2$ continues to rise as $z_2$ decreases through zero.  In
principle, the fit should be symmetric about $z_2 = 0$, but, to
enforce that result it would be necessary to flip the positions of the
other subhalos to the opposite side of the cluster midplane as subhalo
2 crossed the midplane.  This behavior shows that the fit is better
(formally, at the 90\% level) when subhalos 1 and 2 lie on the same
side of the cluster midplane.  Of course, this does not determine
which side of the midplane that is.  Subhalo 3 is too far from the
other two subhalos to obtain any information about its location
relative to them.

When the masses of the subhalos were allowed to vary (Section
\ref{sec:struct}), the best fitting masses for the subhalos associated
with the two gE galaxies near the cluster center 
($m_1 = 9.1\times10^{12}~\msun$ and $m_2 = 7.6\times10^{12}~\msun$) were
found to be consistent with their lensing masses 
($m_{1,{\rm lens}} = (11.0 \pm 1.5) \times10^{12}~\msun$ and 
$m_{2,{\rm lens}} = (6.6 \pm 1.2)\times10^{12}~\msun$),
though not for subhalo 3.  Formally, the large improvements in the fit show that the
X-ray masses determined for the NGC~4889 and NGC~4874 subhalos would
be much more accurate than the lensing masses.  However, given the
shortcomings of our model, particularly the approximation that the
subhalos are static, the results would be dominated by systematic
uncertainties.  More accurate and more realistic models are required to
assess the systematics,  so we make no attempt to pursue this
further here.  The remarkable consistency between the X-ray and
lensing masses for the two central subhalos is noteworthy.  It lends
strong support to the interpretation of these mass concentrations as
subhalos associated with the two gE galaxies, which lie
close to the midplane of the Coma Cluster.  

The lack of any significant amount of cool gas associated with the
subhalos is consistent with these results.  However, more sensitive
observations are needed to detect the temperature enhancements that
should be caused by adiabatic compression associated with a subhalo.
With existing X-ray instruments, the small temperature rise should be
more readily detectable in cooler clusters and groups.  Apart from
providing further constraints on subhalo properties, detecting a
temperature rise will help to distinguish gasless subhalos from other
sources of X-ray enhancements in clusters \citep[e.g.][]{cvz12}.


\section{Conclusions} \label{sec:conc}

We have presented an approximate model for calculating the impact of
initially gasless subhalos on the X-ray image of a cluster which relies on the
subhalos being slow moving and the gas response being adiabatic.  The
model was applied to \xmm{} and \chandra{} data for the Coma Cluster,
using properties determined from lensing data for three central
subhalos reported by \citet{2010Okabe}.  Two of the subhalos correspond to the
gE galaxies, NGC~4889 and NGC~4874.  When combined with an ellipsoidal
model for the gas distribution in the unperturbed cluster, including
the effects of these three subhalos produces a model X-ray image that
is a much better fit to the observed cluster than the ellipsoidal
cluster model alone.

Using the lensing masses for the three subhalos, the distance along
our line of sight of each subhalo from the cluster midplane is tightly
constrained, with 90\% confidence ranges no more than $\pm10$ kpc
($z_1 = 121 \pm 5$ kpc, $z_2 = 7.9 \pm 7.5$ kpc, and $z_3 = 399 \pm 10$ kpc).
The fits are improved considerably when the subhalos associated with
the two gE galaxies are assumed to be centered on those galaxies,
rather than at the lensing positions, which are consistent within the
errors of the lensing determinations.  The model favors the subhalos
associated with the two gE's lying on the same side of the cluster
midplane, further constraining their locations in three dimensions.
The results also show the potential for using X-ray images to constrain
the internal structure of the two central subhalos.

For each of the two gE galaxies, freeing the total mass of the subhalo
gave a significantly better fit ($\Delta\chi^2 \simeq 240$).  For the
NGC~4889 subhalo, the best fitting mass  is
$9.1\times10^{12}~\msun$ \citep[$\simeq 1.2$ standard
deviations  smaller than the lensing mass
of][]{2010Okabe}, while, for NGC~4874, the best fit is
$7.6\times10^{12}~\msun$  ($\simeq 0.9$ standard
deviations more than the lensing mass).  Formally, the X-ray masses are
much better determined, although the results are dominated by
systematic uncertainties that can only be circumvented with more accurate
models.  The X-ray and lensing mass determinations rely on completely
independent data, so that the agreement between them supports the
assumptions that underlie both, most critically that these two subhalos
reside near the center of the Coma Cluster.  Fitting the X-ray image
does not determine a mass for the third subhalo, suggesting that our
model assumptions may be less accurate for it.

Our results highlight the potential for using high quality X-ray data
to probe substructure in galaxy clusters.  Particularly when used in
combination with lensing measurements to constrain the mass of a
subhalo, X-ray data can constrain where along our line of sight a subhalo
lies within a cluster.  The X-ray data can also be used to determine
subhalo masses and, potentially, to probe their internal 
structure, i.e., their truncation and scale radii.

\acknowledgements

We thank R.~Johnson for useful discussions.  The work was supported in
part by Chandra grant AR7-8013X, NASA grant NAS8-03060, 
FAPESP grant 2008/05970-0, and NASA grant NNX11AF76G.


\bibliographystyle{apj}

\input pubs-journals

\bibliography{Coma_dark_matter_nov}

\appendix

\section{Limits of the Approximations}

Two conditions must be satisfied in order for the approximations used
here to be valid.  First, gas displacements caused by a subhalo need
to be small compared to the scale of entropy variations in the
unperturbed atmosphere for the perturbed gas to be treated as
isentropic (Section~\ref{sec:isent}).  Second, for the fluid
acceleration, $d\mathbf{v}/dt$ of Equation~(\ref{eqn:mom}), to be
negligible, it must be small compared to the gravitational
acceleration due to the subhalo, $\nabla\deltaphi$.  Here, we
determine conditions for these approximations to hold.

The first condition will be satisfied if displacements are small
compared to the core radius, $\rc$, of the unperturbed atmosphere
(Equation (\ref{density})).  We can estimate the displacements caused by
a slowly moving subhalo embedded in isentropic gas from the
requirements of mass conservation.  Assuming that displacements around
a subhalo are radial with respect to the center of the subhalo, the
mass of gas within any sphere fixed to the fluid elements of the gas
is fixed.  In the notation of Section \ref{sec:isent}, conservation of
mass therefore requires
\begin{equation}
{4 \pi \over 3} n_0 r^3 = \int_0^{r - \delta r} n(r') 4 \pi r'^2 \, dr',
\end{equation}
where the radial coordinates, $r$ and $r'$, are centered on the
subhalo and $\delta r$ is the inward radial displacement of the gas.
Using Equation (\ref{density_perturbation}) and expanding to first order
in the perturbation, this gives
\begin{equation}
{1\over3} n_0 [r^3 - (r - \delta r)^3] \simeq n_0 r^2 \delta r
\simeq {3 n_0 \over 2 H_0} \int_0^{r} [-\deltaphi(r')] r'^2 \, dr'.
\label{eqn:drest}
\end{equation}

Outside the truncation radius, $\rt$, the gravitational potential for
our subhalo model (Section \ref{sec:subhalomodel}) is simply
$\deltaphi(r) = - G \mtot /r$, with $\mtot$ constant.  Although this
form overestimates $|\delta\Phi|$, hence $\delta r$, for $r < \rt$,
the spherical geometry ensures that the difference it makes to our
estimate for $\delta r$ diminishes rapidly with increasing radius
outside $\rt$.  Using $\deltaphi(r) = - G \mtot / r$ in Equation
(\ref{eqn:drest}) to estimate $\delta r$ (an overestimate), gives
\begin{equation}
\delta r \simeq {3 G\mtot \over 4 H_0},
\label{eqn:dr}
\end{equation}
independent of $r$.  Although the displacements around a moving halo
will not be purely radial, this provides a good estimate for the
magnitude of the actual displacements.  For the subhalos considered
here, $G \mtot / H_0 \ll \rc$ and the first requirement is well
satisfied.

To estimate the fluid accelerations due to a moving subhalo, consider
a frame in which the subhalo remains at rest, centered on the origin.
At large distances from center of the subhalo, the fluid moves through
this frame with velocity $\vvec_0$, where $-\vvec_0$ is the velocity
of the subhalo with respect to the cluster.  In the absence of the
subhalo, the path of a fluid element would be given as a function of
the time by $\rvec_0(t) = \rveci + \vvec_0 t$, where $\rveci$ is a
constant.  Including the approximate radial displacement, $\delta r$,
of Equation~(\ref{eqn:dr}), in the presence of a subhalo the path of the
fluid element is given approximately by
\begin{equation}
\rvec_1(t) = \rvec_0(t) - \delta r {\rvec_0(t) \over |\rvec_0(t)|}.
\end{equation}

Differentiating this with respect to the time once gives the fluid
velocity
\begin{equation}
{d\rvec_1(t) \over dt} = \vvec_0 - {\delta r \over |\rvec_0(t)|^3}
\rvec_0(t) \times [\vvec_0 \times \rvec_0(t)],
\end{equation}
and twice gives the acceleration
\begin{equation}
{d^2 \rvec_1(t) \over dt^2} = 3 {\delta r \over |\rvec_0(t)|^5} \vvec_0
\cdot \rvec_0(t) [\rvec_0(t) \times \{\vvec_0 \times \rvec_0(t)\}] -
      {\delta r \over |\rvec_0(t)|^3} \vvec_0 \times [\vvec_0 \times
        \rvec_0(t)].
\end{equation}

In terms of the unit vectors $\vhat = \vvec_0/|\vvec_0|$ and $\rhat =
\rvec_0(t) / |\rvec_0(t)|$, the cosine of the angle between $\vvec_0$
and $\rvec_0(t)$ is $u = \vhat \cdot \rhat$.  We then have
$|\rvec_0(t)|^2 = b^2 / (1 - u^2)$, where $b$ is the distance of
closest approach of $\rvec_0(t)$ to the center of the dark matter halo
(the impact parameter of the fluid element).  With this notation, the
acceleration can be written as
\begin{equation}
{d^2 \rvec_1(t) \over dt^2} = {\delta r \over b^2} |\vvec_0|^2 (1 -
u^2) \left[ 2 u \vhat + (1 - 3 u^2) \rhat \right].
\end{equation}

Its magnitude is maximized when $u = 0$, \ie, when the fluid is closest
to the center of the subhalo, giving
\begin{equation}
\left|d^2 \rvec_1 \over dt^2\right|_{\rm max} = {\delta r |\vvec_0|^2
  \over b^2} \simeq {3 G\mtot |\vvec_0|^2 \over 4 H_0 b^2 }.
\end{equation}

The gravitational acceleration due to the subhalo at this point is
approximately $G\mtot / b^2$, so the requirement that the fluid
acceleration be small compared to the gravitational acceleration is
simply 
\begin{equation}
|\vvec_0|^2 \ll {4\over 3} H_0 = 2 s_0^2,
\end{equation}
where we have expressed the specific enthalpy in terms of the squared
sound speed ($H_0 = 3s_0^2/2$ for $\gamma = 5/3$).  This requirement
imposes the condition that the speed of the subhalo should be
transonic at most.

\end{document}

%% file: pubs-journals.tex



\newcommand\jcap{JCAP}